\documentclass[trackchanges,twocolumn]{aastex701}

\usepackage{multirow}

\begin{document}

\title{JWST Spectroscopy of GRB\,250702B: An Extremely Rare and Exceptionally Energetic Burst in a Dusty, Massive Galaxy at $z=1.036$}

\author[orcid=0000-0002-5826-0548]{Benjamin P. Gompertz}
\affiliation{School of Physics and Astronomy, University of Birmingham, Birmingham B15 2TT, UK}
\affiliation{Institute for Gravitational Wave Astronomy, University of Birmingham, Birmingham B15 2TT, UK}
\email[show]{b.gompertz@bham.ac.uk}  

\author[orcid=0000-0001-7821-9369]{Andrew J. Levan}
\affiliation{Department of Astrophysics/IMAPP, Radboud University, P.O. Box 9010, 6500 GL Nijmegen, The Netherlands}
\email{}

\author[0000-0003-1792-2338]{Tanmoy Laskar}
\affiliation{Department of Physics \& Astronomy, University of Utah, Salt Lake City, UT 84112, USA}
\affiliation{Department of Astrophysics/IMAPP, Radboud University, P.O. Box 9010, 6500 GL Nijmegen, The Netherlands}
\email{}

\author[0000-0003-4876-7756]{Benjamin Schneider}
\affiliation{Aix Marseille Univ., CNRS, CNES, LAM, Marseille, France}
\email{}

\author[0000-0001-9842-6808]{Ashley A. Chrimes}
\affiliation{European Space Agency (ESA), European Space Research and Technology Centre (ESTEC), Keplerlaan 1, 2201 AZ Noordwijk, the Netherlands}
\affiliation{Department of Astrophysics/IMAPP, Radboud University, P.O. Box 9010, 6500 GL, Nijmegen, The Netherlands}
\email{}

\author[orcid=0000-0001-5108-0627]{Antonio Martin-Carrillo}
\affiliation{School of Physics and Centre for Space Research, University College Dublin, Belfield, Dublin 4, Ireland}
\email[]{antonio.martin-carrillo@ucd.ie}  

\author[0000-0002-5460-6126]{Albert Sneppen}
\affiliation{Niels Bohr Institute, University of Copenhagen, Jagtvej 155, 2200, Copenhagen N, Denmark}
\affiliation{The Cosmic Dawn Centre (DAWN), Denmark}
\email{}

\author[orcid=0009-0001-1554-1868]{David O’Neill}
\affiliation{School of Physics and Astronomy, University of Birmingham, Birmingham B15 2TT, UK}
\affiliation{Institute for Gravitational Wave Astronomy, University of Birmingham, Birmingham B15 2TT, UK}
\email[]{d.s.oneill@bham.ac.uk}

\author[0000-0002-7517-326X]{Daniele B. Malesani} 
\affiliation{Niels Bohr Institute, University of Copenhagen, Jagtvej 155, 2200, Copenhagen N, Denmark}
\affiliation{The Cosmic Dawn Centre (DAWN), Denmark}
\affiliation{Department of Astrophysics/IMAPP, Radboud University, P.O. Box 9010, 6500 GL, Nijmegen, The Netherlands}
\email{}

\author[orcid=0000-0001-5679-0695, gname=Peter, sname=Jonker]{Peter G. Jonker}
\affiliation{Department of Astrophysics/IMAPP, Radboud University, P.O. Box 9010, 6500 GL, Nijmegen, The Netherlands}
\email[]{p.jonker@astro.ru.nl}

\author[0000-0002-2942-3379]{Eric~Burns}
\affiliation{Department of Physics \& Astronomy, Louisiana State University, Baton Rouge, LA 70803, USA}
\email{ericburns@lsu.edu}

\author[orcid=0009-0009-1573-8300]{Gregory Corcoran}
\affiliation{School of Physics and Centre for Space Research, University College Dublin, Belfield, Dublin 4, Ireland}
\email[]{gregory.corcoran@ucdconnect.ie}  

\author[orcid=0000-0002-7910-6646]{Laura Cotter}
\affiliation{School of Physics and Centre for Space Research, University College Dublin, Belfield, Dublin 4, Ireland}
\email[]{laura.cotter@ucdconnect.ie}

\author[orcid=0000-0001-7717-5085, gname=Antonio, sname=de Ugarte Postigo]{Antonio de Ugarte Postigo}
\affiliation{Aix Marseille Univ., CNRS, CNES, LAM, Marseille, France}
\email[]{antonio.deugarte@lam.fr}

\author[0000-0001-9868-9042]{Dimple}
\affiliation{School of Physics and Astronomy, University of Birmingham, Birmingham B15 2TT, UK}
\affiliation{Institute for Gravitational Wave Astronomy, University of Birmingham, Birmingham B15 2TT, UK}
\email{d.dimple@bham.ac.uk}

1   §\author[orcid=0000-0002-8775-2365]{Rob A. J. Eyles-Ferris}
\affiliation{School of Physics and Astronomy, University of Leicester, University Road, Leicester, LE1 7RH, UK}
\email[]{raje1@leicester.ac.uk}  

\author[orcid=0000-0001-9695-8472]{Luca Izzo}
\affiliation{INAF, Osservatorio Astronomico di Capodimonte, Salita Moiariello 16, I-80121 Naples, Italy}
\affiliation{DARK, Niels Bohr Institute, University of Copenhagen, Jagtvej 155A, 2200 Copenhagen, Denmark}
\email[]{luca.izzo@inaf.it} 

\author[orcid=0000-0002-9404-5650]{P\'all Jakobsson}
\affiliation{Centre for Astrophysics and Cosmology, Science Institute, University of Iceland, Dunhagi 5, 107 Reykjavík, Iceland}
\email[]{pja@hi.is}  

\author[0000-0001-5169-4143]{Gavin P. Lamb}
\affiliation{Astrophysics Research Institute, Liverpool John Moores University, IC2 Liverpool Science Park, 146 Brownlow Hill, Liverpool, L3 5RF, UK}
\email[]{g.p.lamb@ljmu.ac.uk}

\author[orcid=0000-0002-9408-1563]{Jesse T. Palmerio}
\affiliation{Universit\'e Paris-Saclay, Universit\'e Paris Cit\'e, CEA, CNRS, AIM, 91191, Gif-sur-Yvette, France}
\email[]{jesse.palmerio@cea.fr}  

\author[orcid=0000-0003-3457-9375, gname=Giovanna, sname=Pugliese]{Giovanna Pugliese}
\affiliation{Anton Pannekoek Institute of Astronomy, University of Amsterdam, Science Park 904, 1098 XH Amsterdam, The Netherlands}
\email[]{pugliese@astroduo.org}

\author[orcid=0000-0003-3193-4714, gname=Maria Edvige, sname=Ravasio]{Maria Edvige Ravasio}
\affiliation{Department of Astrophysics/IMAPP, Radboud University, P.O. Box 9010, 6500 GL, Nijmegen, The Netherlands}
\affiliation{INAF-Osservatorio Astronomico di Brera, Via Bianchi 46, I-23807, Merate (LC), Italy}
\email[]{mariaedvige.ravasio@ru.nl}

\author[0000-0002-6950-4587]{Andrea Saccardi}
\affiliation{Universit\'e Paris-Saclay, Universit\'e Paris Cit\'e, CEA, CNRS, AIM, 91191, Gif-sur-Yvette, France}
\email[]{andrea.saccardi@cea.fr}

\author[orcid=0000-0002-9393-8078]{Ruben Salvaterra}
\affiliation{INAF-Istituto di Astrofisica Spaziale e Fisica Cosmica di Milano, Via A. Corti 12, 20133 Milano, Italy}
\email[]{ruben.salvaterra@inaf.it}  

\author[0000-0003-2700-1030]{Nikhil Sarin}
\affiliation{Kavli Institute for Cosmology, University of Cambridge, Madingley Road, CB3 0HA, UK}
\affiliation{Institute of Astronomy, University of Cambridge, Madingley Road, CB3 0HA, UK}
\email{nikhil.sarin@ast.cam.ac.uk}

\author[orcid=0000-0001-6797-1889, gname=Steve, sname=Schulze]{Steve Schulze}
\affiliation{Center for Interdisciplinary Exploration and Research in Astrophysics (CIERA), Northwestern University, 1800 Sherman Ave., Evanston, IL 60201, USA}
\email[]{steve.schulze@northwestern.edu}

\author[orcid=0000-0003-3274-6336, gname=Nial, sname=Tanvir]{Nial Tanvir}
\affiliation{School of Physics and Astronomy, University of Leicester, University Road, Leicester, LE1 7RH, UK}
\email[]{nrt3@le.ac.uk}

\author[0009-0009-8473-3407]{Makenzie E. Wortley}
\affiliation{School of Physics and Astronomy, University of Birmingham, Birmingham B15 2TT, UK}
\affiliation{Institute for Gravitational Wave Astronomy, University of Birmingham, Birmingham B15 2TT, UK}
\email[]{mwortley@star.sr.bham.ac.uk}

\begin{abstract}

We present follow-up observations of the day-long, repeating GRB\,250702B with the Near Infrared Spectrograph (NIRSpec) on board the James Webb Space Telescope (JWST). Through the identification of narrow hydrogen emission lines at a consistent redshift of $z = 1.036 \pm 0.004$, we calibrate the distance scale, and therefore the energetics, of this unique extragalactic transient. At this distance, the resulting $\gamma$-ray energy release is at least $E_{\gamma,\rm iso} = 2.2 \times 10^{54}$\,erg. We find no evidence for ongoing transient emission at the GRB position, and exclude any accompanying supernova with a luminosity comparable to the Type Ic broad-line SN\,2023lcr, though we are unable to rule out a fainter SN counterpart due to high extinction. The inferred rate of such events, assuming at most one in the lifetime of {\em Fermi}, suggests that such bursts are very rare, with volumetric rates over $1000$ times lower than normal high-luminosity long GRBs and $> 10^5$ times lower than core collapse supernovae, when corrected for beaming. Furthermore, we find that the host galaxy is unique amongst GRB host galaxies, and extremely rare in the general galaxy population, being extremely large, dusty, and with high stellar mass. The identification of such an exotic GRB in such an unusual galaxy raises the possibility that the environment was important in the progenitor channel creating GRB 250702B. 

\end{abstract}

\keywords{\uat{Galaxies}{573} --- \uat{High Energy astrophysics}{739}}


\section{Introduction} 

Gamma-ray bursts (GRBs) are extremely high-energy explosive transients that are generally thought produced by processes internal to relativistic jets launched by either the collapse of very massive stars at the ends of their lives, or by the merging of compact object binaries, predominantly (or possibly exclusively) neutron stars. Their gamma-ray emission is observed to be remarkably diverse, perhaps reflecting the disparate properties and environments of their stellar progenitors combined with the complex physics of jet launch, propagation, and radiation. Broadly, GRBs with durations of greater than 2 seconds are believed to originate from stellar collapse (termed `collapsars') thanks to their consistent association with broad-lined stripped envelope (Type Ic-BL) supernovae \citep[SNe; e.g.][]{Hjorth03,Stanek03,Cano2017}, though notable exceptions to this duration-based dichotomy have emerged \citep{Fynbo06,Gehrels06,Gal-Yam06,Rastinejad22,Troja22,Gompertz23,Levan24,Yang24}. Collapsar GRBs have typical durations of tens to hundreds of seconds in gamma-ray emission, though rare examples of `ultra-long' GRBs with gamma-ray durations of thousands to tens of thousands of seconds have been observed \citep{Gendre13,Levan14}.

Despite this diversity, the recent identification of GRB\,250702B as an extragalactic transient \citep{Levan25} has confirmed the burst as a truly singular event in the GRB population. Unlike all previous examples, GRB\,250702B exhibited repeating structure, triggering the \emph{Fermi} Gamma-ray Burst Monitor \citep[GBM;][]{Meegan09} three times on 2025-07-02 \citep[initially identified as GRBs 250702B, D, and E;][]{gcn40931}, while prior X-ray emission from the same position was reported by the \emph{Einstein Probe} \citep[EP;][]{gcn40906} starting on 2025-07-01. GRBs have never been seen to repeat on these timescales, and the isolated `islands' of gamma-ray emission are unlike anything previously seen in even the ultra-long class of burst \citep[though separations of spikes on the order of kiloseconds have been observed, e.g.][]{Virgili13}.

Indeed, due to the peculiarity
of the burst, the off-nuclear position of the transient on its host galaxy, and the identification of a possible periodicity in the gamma-ray triggers, \citet{Levan25} suggested GRB\,250702B may have instead arisen from the tidal disruption event (TDE) of a white dwarf around an intermediate mass black hole (IMBH). A TDE origin was also proposed by \citet{Oganesyan25}. Jetted, or relativistic, TDEs have previously been observed with properties similar to GRBs: an initial strong burst of gamma-rays (though with durations far in excess of those typical for GRBs), followed by a fading, multi-wavelength X-ray-to-radio counterpart \citep{Burrows11, Cenko12, Pasham15, Levan16, Mangano16}. The spectral energy distributions of relativistic TDEs have also shown them to be powered by a combination of thermal and synchrotron emission, and are akin to those of GRBs \citep{Andreoni22,Pasham23}. In their modeling, \citet{Levan25} concluded that the X-ray, near-infrared (nIR) and radio afterglow following GRB\,250702B was consistent with more typical GRBs if the redshift of the source was $z \sim 0.3$, but may be more consistent with the luminosity (if not the decay rate) of relativistic TDEs at higher $z$. This underscores the importance of having a measured distance with which to estimate the energy release and evaluate potential progenitors.

Here, we report on follow-up observations of GRB\,250702B with the NIRSpec spectrograph on board the James Webb Space Telescope (JWST). We measure the redshift of the burst to be $z = 1.036 \pm 0.004$, and discuss the resultant implications on the energetics. We also provide limits on an accompanying SN Ic-BL, which would be expected for a collapsar progenitor, and discuss our constraints on TDE progenitor. Finally, we analyse the properties of the host galaxy of GRB\,250702B and discuss how this contextualises the transient. This letter is organised as follows: In Section~\ref{sec:obs_z} we outline our observations and derive a redshift for the burst. We place limits on transient counterparts in Section~\ref{sec:transients}. The properties of the host galaxy are presented in Section~\ref{sec:galaxy}. We discuss the implications of our findings in Section~\ref{sec:discussion} and summarise our final conclusions in Section~\ref{sec:conclusions}. We assume a cosmology of $H_0 = 69.6$\,km\,s$^{-1}$, $\Omega_M = 0.286$, and $\Omega_{vac} = 0.714$ \citep{Bennett14} throughout.

\section{Observations and redshift determination} \label{sec:obs_z}

We observed the position of the nIR counterpart to GRB\,250702B \citep{Levan25} at 03:12:28 UTC on 2025-08-23, 51.6 days after the \emph{Fermi} trigger GRB~250702D under program GO 9254 (PI: Gompertz) with the NIRSpec unfiltered prism and 0.4$\arcsec$ S400A1 fixed slit for an effective exposure time of 1.7\,hours. We utilize the standard level 3 pipeline products from the Mikulski Archive for Space Telescopes (MAST) in our analysis.

We also observed GRB\,250702B with the Enhanced Resolution Imager and Spectrograph (ERIS) mounted on Unit Telescope 4 (Yepun) of the Very Large Telescope (VLT) at Cerro Paranal Observatory, Chile. Observations were taken with the Near Infrared Camera System (NIX) starting at 00:46:25 UTC on 2025-08-23, (51.5 days after the \emph{Fermi}/GBM trigger) under program 114.27PZ (PIs: Malesani, Vergani, Tanvir) with the Ks filter.
The ERIS/NIX images were processed using the ESO \textsc{Reflex} pipeline \citep{Freudling13}, which applies detector-level calibrations, flat-fielding, and sky subtraction. From the individual reduced frames, we performed our own stacking to improve the background subtraction and enhanced the final image quality.

\subsection{1D spectral analysis}
\label{sec:1d_spec}
\begin{figure*}
\centering
    \includegraphics[width=16cm]{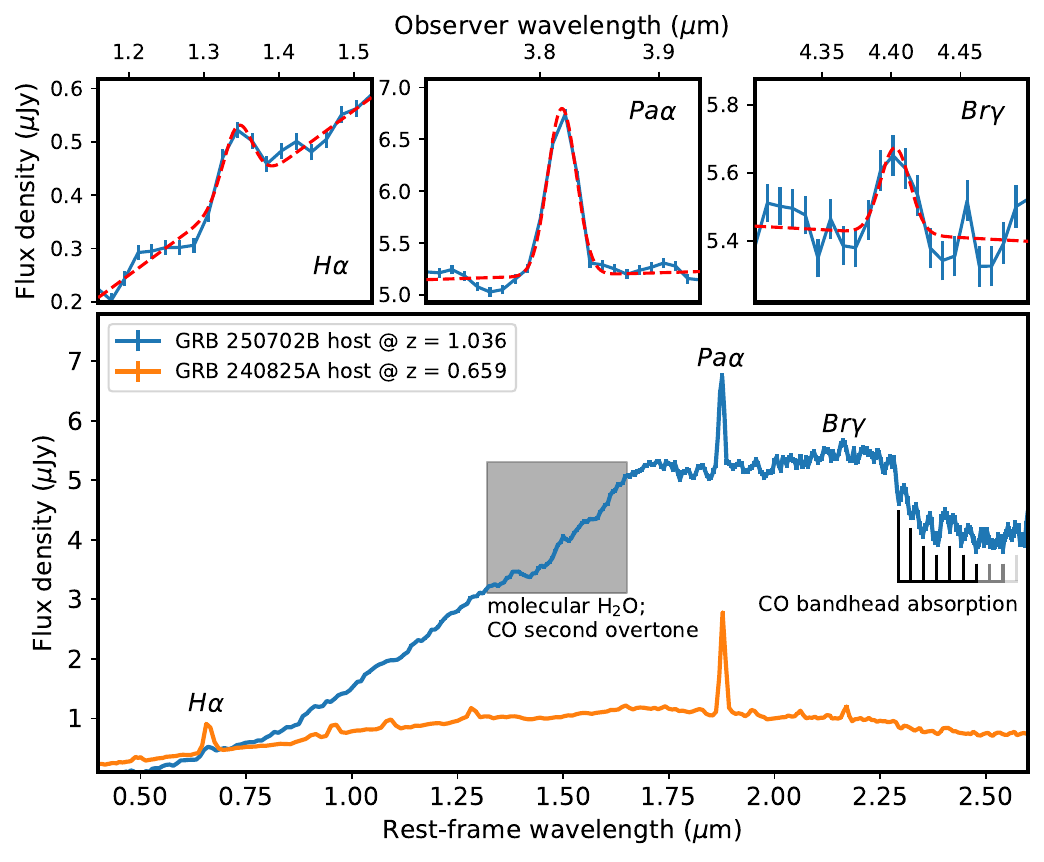}
    \caption{Rest-frame 1D spectrum (lower panel) of GRB\,250702B (blue) at the best-fit redshift of $z = 1.036$, compared to an off-transient (host-galaxy) spectrum of GRB 240825A (orange) at $z = 0.659$ \citep{Martin-Carrillo24}. The high significance Pa$\alpha$, moderate significance H$\alpha$, and low significance Br$\gamma$ lines are marked and their fits are shown in the top panels.  A sharp dropoff in flux is seen redwards of $\sim 2.25$\,$\mu$m in the rest frame, likely due to CO bandhead absorption. The expected wavelengths of individual bandhead components (CO 2–0 through CO 11–9) are indicated in progressively lighter shades. Broad absorption features are also identified between $1.3$\,$\mu$m and $1.6$\,$\mu$m, consistent with expectations of absorption from molecular H$_2$O and second-overtone CO.}
    \label{fig:1d_spec}
\end{figure*}

We de-redden the 1D spectrum for a Milky Way extinction of $A_V = 0.83$ along the line of sight \citep{Schlafly11} using the {\sc extinction} Python package\footnote{\url{https://extinction.readthedocs.io/en/latest/index.html}} and a \citet{Cardelli89} extinction law with $R_v = 3.1$. The corrected spectrum still exhibits a very red continuum\footnote{Fitting the $1.5$--$2.5\,\mu$m region with a power-law model yields $\beta_\lambda\approx3$ (for $F_\lambda\propto\lambda^\beta$), corresponding to $F_\nu\propto\nu^{-5}$.}; this is consistent with the high host $A_V$ inferred by \citet{Levan25}, although their solution was for an assumed $z \approx 0.16$. A single high-significance emission line is seen at a wavelength of $\lambda_{\rm obs} \approx 3.8$\,$\mu$m, along with other lower significance line candidates, most notably at $\lambda_{\rm obs} \approx 1.3$\,$\mu$m. Due to the paucity of available lines, we employ template matching to constrain the possible space of redshift solutions. Comparing with the NIRSpec spectrum of the host galaxy of GRB 240825A\footnote{Also taken under program GO 9254. We extract a spectrum at a position offset from the GRB and deredden it using the same method as above but for $A_V = 0.165$ \citep{Schlafly11}. A full analysis of this GRB will be published in Schneider et al. (2025, in prep).} \citep[$z = 0.659$;][]{Martin-Carrillo24}, we identify the high-significance line at $\lambda_{\rm obs}\approx3.8\,\mu$m in the JWST spectrum of GRB\,250702B to be most likely due to Pa$\alpha$ at a redshift of $z \approx 1.03$. At this redshift, the $1.3$\,$\mu$m line is consistent with H$\alpha$ and the continuum shape also appears congruent with that of the template. In addition, by comparison with GRB\,240825A, we identify a possible emission line at $\lambda_{\rm obs}\approx 4.4$\,$\mu$m, which would be consistent with Br$\gamma$ at $z \approx 1.03$. The rest frame spectrum of GRB\,250702B along with the GRB 240825A template is shown in Figure~\ref{fig:1d_spec} with the above line identifications marked. We note a very strong dropoff in flux at the red end of the spectrum, starting at around $4.5$\,$\mu$m in the observer frame. Numerous low-significance features are visible in this region at wavelengths that closely match expectations for the CO first-overtone absorption bandhead \citep{Marmol-Queralto08}. We also identify possible broad absorption features in the region between $1.3$\,$\mu$m and $1.6$\,$\mu$m in the rest frame, consistent with expectations for molecular H$_2$O absorption \citep[e.g.][]{Lancon00} and second-overtone CO absorption \citep[e.g.][]{Tenenbaum05}. These are marked in Figure~\ref{fig:1d_spec}. The implications of these features are discussed in Section~\ref{sec:galaxy}.

To constrain the redshift as well as the strength and significance of the narrow spectral lines more precisely, we fit the line candidates with Gaussian profiles. First, we isolate arbitrary wavelength regions around the line candidates and fit each with a simple Gaussian to estimate line centroid locations ($m$) and widths ($\sigma$). We then extract a region of 10$\sigma$ on either side of the line centroid and fit each line with a Gaussian + 1D polynomial model to separate out the underlying continuum using the {\sc astropy.modeling} package and the Levenberg-Marquardt least squares fitting algorithm. The fit results are presented in Table~\ref{tab:lines}. Line detection significances are given as the ratio of the integrated fluxes to their corresponding measurement errors. Fitting the derived line centroids to their expected rest values in vacuum yields a redshift of $z = 1.036 \pm 0.001$ (however, see Section~\ref{sec:2d_spec} for an additional source of uncertainty in this measurement).
Assuming an intrinsic Case B line ratio of Pa$\alpha$ to H$\alpha$ of 0.122 (corresponding to an electron temperature, $T_e\approx10^4$K and density, $n_e\approx10^{-2}$--$10^{4}$\,cm$^{-3}$) along with $R_v = 3.1$ and a \citet{Cardelli89} extinction law, the observed fluxes of these two line candidates in the integrated spectrum imply an intrinsic extinction of $A_V=3.4\pm0.4$~mag from the host galaxy. We note that unmodeled contribution to the H$\alpha$ line (e.g., from blended NII emission) would result in the derived $A_V$ values reported here to be underestimated. 

\begin{table*}[]
    \centering
    \begin{tabular}{ccccccc}
    \hline\hline
    Feature & $\lambda_{\rm obs}$ & $\lambda_{\rm rest}$ & $\sigma$ & Integrated flux (fit) & Integrated flux (fixed) & S/N \\
     &  ($\mu$m) & ($\mu$m) & (nm) & ($10^{-18}$\,erg\,s$^{-1}$\,cm$^{-2}$) & ($10^{-18}$\,erg\,s$^{-1}$\,cm$^{-2}$) & \\
    \hline
    H$\alpha$ & $1.345 \pm 0.003$ & $0.656$ & $19.5 \pm 3.3$ & $10.5 \pm 2.30$ & $6.48 \pm 0.97$ & $4.5$ \\
    Pa$\alpha$ & $3.818 \pm 0.001$ & $1.875$ & $12.0 \pm 0.7$ & $9.94 \pm 0.75$ & $9.94 \pm 0.75$ & $13.2$ \\
    Br$\gamma$ & $4.403 \pm 0.003$ & $2.166$ & $10.5 \pm 3.2$ & $1.01 \pm 0.41$ & $1.16 \pm 0.31$ & $2.5$ \\
    \hline\hline
    \end{tabular}
    \caption{Line identifications and Gaussian fit results. Integrated fluxes (corrected for Galactic extinction) are presented for the value of $\sigma$ found for each line (fit), and with all line widths fixed to the Pa$\alpha$ value (fixed) to minimize flux uncertainties in the less significant lines. The identified lines are consistent with a redshift of $z = 1.036 \pm 0.001$.}
    \label{tab:lines}
\end{table*}

\subsection{2D spectral analysis}\label{sec:2d_spec}
\begin{figure*}
\centering
    \includegraphics[width=16cm]{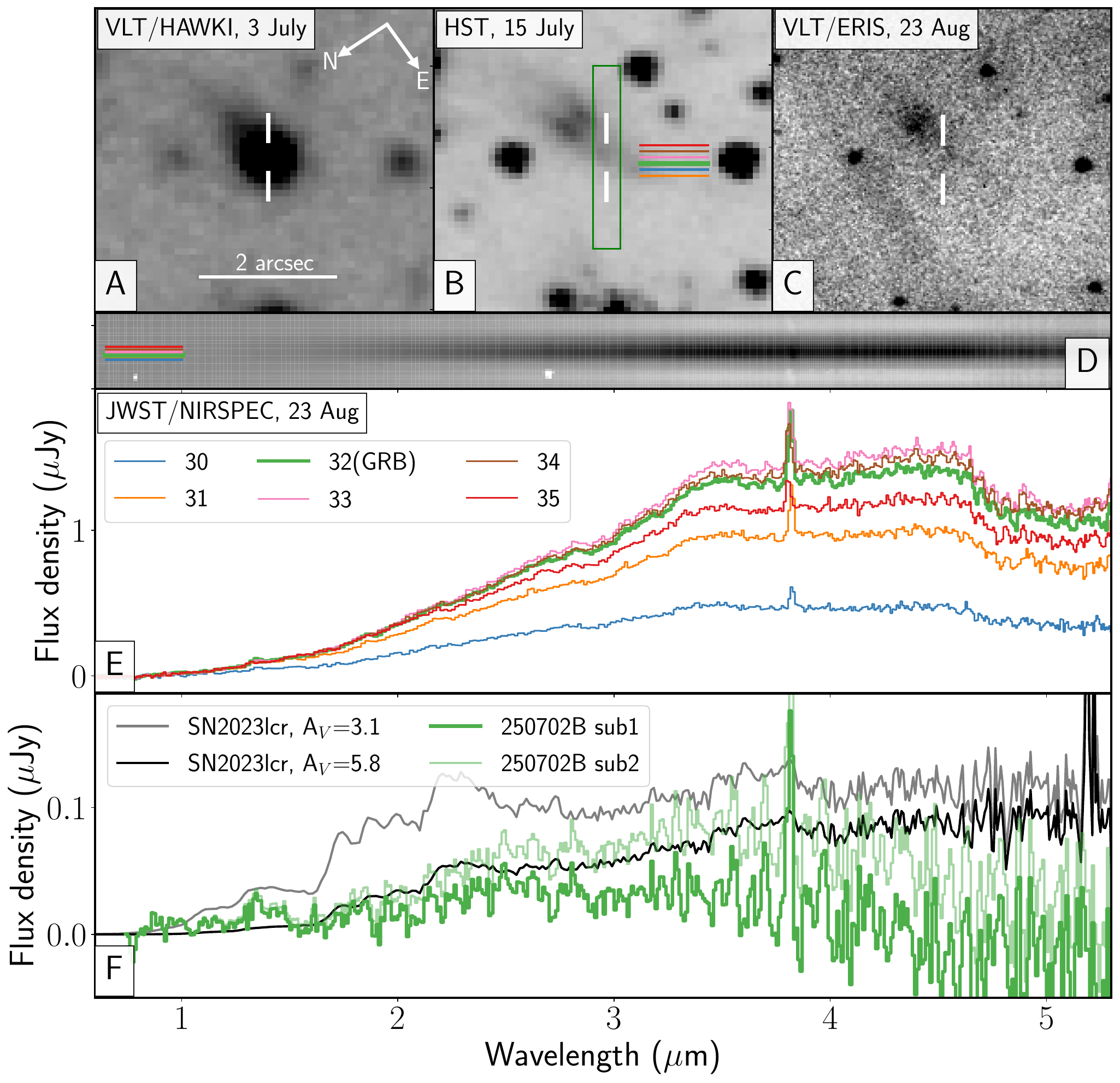}
    \caption{Spatially resolved properties of the GRB 250702B host galaxy, and supernova limits. Panels A-C show imaging of the source location from July to August using the VLT and HST \citep{Levan25}. The location of the slit is shown in panel B along with lines indicating the physical locations that correspond to each row in the extracted spectrum. The row containing the GRB is shown in green (row 32), and the resulting 2D spectrum is shown in panel D. Panel E shows the 1D spectrum (not corrected for host or Galactic extinction) extracted from each row. The spectral shape is generally very similar in each row with no evidence for additional emission components at the GRB location. We demonstrate this in panel F where we subtract a host galaxy spectrum (row 34) from that at the GRB location. Depending on the normalisation of the subtraction it is possible to have near zero flux (sub1), or a faint component that is essentially a copy of the host galaxy spectrum (sub2). We also plot SN\,2023lcr as it would appear with $A_V=3.1$ and $A_V=5.7$, corresponding to the inferred extinction at the burst location from the row-by-row spectral extraction (Table~\ref{tab:per-pix-linefluxes}) and afterglow modeling (Section~\ref{sec:afterglow}), respectively. In the lower extinction scenario the absence of broad features can exclude supernova emission to factors of 2-3 fainter than SN\,2023lcr. In the higher extinction scenario it is not possible to place meaningful constraints.}
    \label{fig:pixelspec}
\end{figure*}

To isolate potential light from the transient from that of the underlying galaxy, we perform a spatially resolved study by extracting spectra along individual rows near the center of the slit (pixel rows 30--35) with sufficient signal via the {\tt extract1d} too in the JWST pipeline. We show the resulting output, along with its correspondence to the host,  in Figure~\ref{fig:pixelspec}. We correct each of these spectra for Galactic extinction as above, and fit the three atomic H spectral features identified above (H$\alpha$, Pa$\alpha$, and Br$\gamma$) with Gaussian models using a similar procedure as in Section~\ref{sec:1d_spec}. We fix the width of all three features to be the same in a given spectrum, but allow the fluxes (and the underlying continuum) as well as the redshift, to vary across the six spectra. In each case, we use the flux ratio of Pa$\alpha$ to H$\alpha$ to estimate $A_V$. Finally, we convert the inferred redshift from each spectrum to a velocity relative to $z=1.036$ inferred from the integrated spectrum (Section~\ref{sec:1d_spec}) and present the results in Table~\ref{tab:per-pix-linefluxes}.  

\begin{table*}[t]
    \centering
    \begin{tabular}{cccccccc}
    \hline\hline
    Pixel & $\delta z$ & $v$ & $\sigma$ & $F_{\rm Pa\alpha}$ & $F_{\rm H\alpha}$ & $F_{\rm Br\gamma}$ & $A_V$\\
    Row & ($10^{-4}$)  & (km s$^{-1}$) & (nm) & \multicolumn{2}{c}{($10^{-18}$\,erg\,s$^{-1}$\,cm$^{-2}$)} &     ($10^{-19}$\,erg\,s$^{-1}$\,cm$^{-2}$) & (mag)
    \\
    \hline
    30 & $28.2\pm7.6$ & $850\pm230$ & $11.2\pm2.0$ & $0.85\pm0.13$ & $1.22\pm0.36$ & $0.60\pm0.50$ & $4.00\pm0.52$ \\
    31 & $18.6 \pm 3.5$ & $560\pm100$ & $10.7 \pm 0.8$ & $2.02 \pm 0.14$ & $1.99 \pm 0.33$ & $1.78 \pm 0.80$ & $3.38\pm0.29$\\
    32 & $-2.0 \pm 3.0$ & $-60\pm90$ & $11.8 \pm 0.6$ & $3.15 \pm 0.17$ & $2.54 \pm 0.35$ & $1.39 \pm 0.84$ & $3.07\pm0.23$\\
    33 & $-22.6 \pm 3.5$ & $-680\pm100$ & $12.4 \pm 0.7$ & $3.01 \pm 0.19$ & $1.87 \pm 0.35$ & $3.34 \pm 1.10$ & $2.66\pm0.32$ \\
    34 & $-59.1 \pm 4.2$ & $-1780\pm130$ & $11.7 \pm 1.0$ & $2.22 \pm 0.19$ & $1.59 \pm 0.35$ & $3.74 \pm 1.08$ & $2.87\pm0.39$ \\
    35 & $-69.4 \pm 6.6$ & $-2400\pm150$ & $10.7 \pm 1.6$ & $1.19 \pm 0.17$ & $1.08 \pm 0.34$ & $0.73 \pm 0.62$ & $3.26\pm0.61$ \\ 
    \hline\hline
    \end{tabular}
    \caption{Results from fits to spectra extracted row-by-row (Section~\ref{sec:2d_spec}). Pixel numbers correspond to the spectra in Figure~\ref{fig:pixelspec}. The redshift offset $\delta z$ is computed with respect to the systemic redshift derived from the integrated spectrum, $z=1.036$ and velocities are $v=c\delta z$. Line fluxes are corrected for Galactic extinction but not for host extinction. $A_V$ calculated from Case B recombination assuming intrinsic ratio of ${\rm Pa}\alpha/{\rm H}\alpha = 0.122$ and \citet{Cardelli89} dust law with $R_V=3.1$. The centroid of transient emission is expected in pixel row 32 (Section~\ref{sec:2d_spec}).}
    \label{tab:per-pix-linefluxes}
\end{table*}

The inferred central velocity of the spectral lines shifts monotonically from $\approx850$\,km\,s$^{-1}$ to $\approx-2400$\,km\,s$^{-1}$ over the six pixel rows. At a redshift of $z=1.036$, the NIRSpec pixel scale of $0.1\arcsec$/pixel corresponds to a spatial resolution of $0.82$\,kpc/pix. If real, this velocity structure would imply an improbably large enclosed mass of $M\approx10^{12}(v/10^3\,{\rm km\,s}^{-1})^2(R/4.9\,{\rm kpc})M_\odot$. 
Instead, we suggest this signature to be the result of a spatially extended emission feature in the galaxy that is misaligned with the spatial direction of the slit (see Figure~\ref{fig:pixelspec}). Since the width of the NIRSpec point-spread function ($\approx0.1\arcsec$) is less than the width of the employed slit (S400A1; $0.4\arcsec$), such a misalignment would cause a shift in the line centroid along the dispersion direction. This correlation between source placement in the slit and inferred velocity is an additional source of instrumental systematic in the redshift measurement. To account for this effect, we include the standard deviation of the redshift measurements in the per-pixel spectra ($\sigma_z = 0.004$) in quadrature with our redshift uncertainty, giving a final redshift of $z = 1.036 \pm 0.004$. 

We similarly note trends in the derived $A_V$ values along the spatial direction, although the standard deviation of the six $A_V$ values ($\sigma_{\rm A_V}=0.4$\,mag) is consistent with the mean measurement uncertainty ($\bar{\sigma}_{A_V}=0.4$). The mean $A_V=3.1$ across the six pixels is consistent at $<1\sigma$ with $A_V$ derived from the integrated light spectrum (Section~\ref{sec:1d_spec}). Thus, while there is a hint that $A_V$ varies across the galaxy, the effect does not appear to be statistically significant. 

\section{Transient properties and limits}\label{sec:transients}

\subsection{Afterglow modeling}
\label{sec:afterglow}
\begin{figure*}
\centering
    \includegraphics[width=16cm]{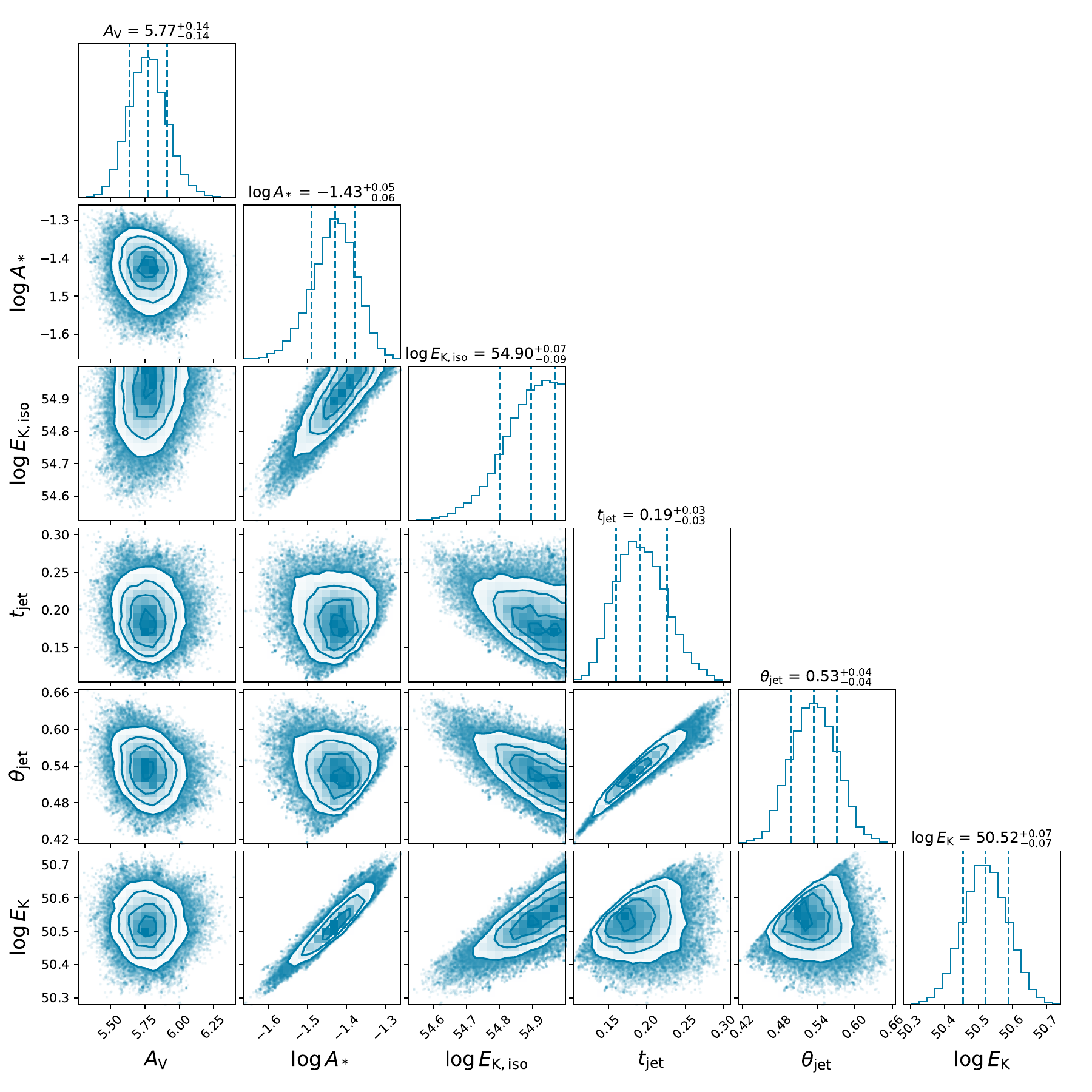}
    \caption{Results from modeling the multi-wavelength X-ray, optical, and radio observations of GRB\,250702B reported in \cite{Levan25} with a standard GRB afterglow model.}
    \label{fig:afterglow-corner}
\end{figure*}

We investigate the energetics of the multi-wavelength counterpart of GRB\,250702B by updating the afterglow model of \cite{Levan25}, utilizing the same dataset but now fixing $z=1.036$. We run 512 Markov Chain Monte Carlo (MCMC) chains for 2000 steps using \textsc{emcee} \citep{emcee2013}, discarding the first 20 steps as burn-in. Our best-fit light curves are visually indistinguishable from those in \citet{Levan25} and are not plotted here, but we summarize our results in Figure~\ref{fig:afterglow-corner}. At this redshift, we infer a very large afterglow isotropic-equivalent kinetic energy of $E_{\rm K,iso}\approx8\times10^{54}$~erg and acknowledge that this parameter is up against our prior of $E_{\rm K,iso}<10^{55}$~erg. However, we stress that the paucity of constraints from the data (e.g., unobserved synchrotron self-absorption and cooling breaks; \citealt{Levan25}) imply degeneracies in this model (Figure~\ref{fig:afterglow-corner}). An anticorrelation between $E_{\rm K,iso}$ and the jet-break time ($t_{\rm jet}$) in this model results in a somewhat constrained, very narrow opening angle of $\theta_{\rm jet}\approx0.5^\circ$, similar to the value derived in \citet{Levan25}. The corresponding beaming-corrected kinetic energy of $E_{\rm K}\approx3\times10^{50}$~erg remains consistent with typical values inferred for long-duration GRBs (e.g., \citealt{laskar2015}). 
This model requires $\approx5.8$\,mag of visual extinction along the line-of-sight to the transient. Although lower\footnote{The lower $A_V$ is expected at the higher redshift employed here, since the rest-frame wavelengths sampled by the observed NIR bands are correspondingly bluer.} than the value ($A_V\approx11.6$\,mag) reported in \citet{Levan25}, this relatively high value  is nevertheless demanded\footnote{From the VLT observations at $\delta t\approx1.6$\,days, $H-K\approx1.5$\,AB mag. For a MW extinction curve at $z=1.036$, this constrains $A_V\approx(6.8+1.4\beta)\,{\rm mag}\approx5.9$\,mag with $\beta\sim(1-p)/2\approx-0.55$ for $p\approx2.2$, consistent with the estimates from our model (Figure~\ref{fig:afterglow-corner}).} by the very red nIR color of the afterglow. This $A_V$ is higher than the host-averaged $A_V$ inferred from the JWST spectrum ($A_V=3.4\pm0.4$\,mag; Section~\ref{sec:obs_z}), but such a disparity is not atypical for long-duration GRBs (e.g., \citealt{Schroeder2022}). Finally, we confirm that there is negligible ($\lesssim1\%$) contribution at 51.6~days from the afterglow to the observed JWST spectrum at all wavelengths in this model.

\subsection{Supernova limits}\label{sec:supernova}
In the standard collapsar scenario, we expect the emergence of a Type Ic-BL SN, peaking on a timescale of $\sim2$\,weeks in the rest frame (e.g., \citealt{Cano2017}). The detection or exclusion of an SN therefore has valuable implications for the progenitor of GRB\,250702B. At the derived redshift of $z = 1.036$, our NIRSpec observations were taken around 25.5 rest-frame days after collapse, or around 11 rest-frame days after the expected peak of an accompanying SN.

We use SN\,2023lcr as the template for our expected SN counterpart. This has several advantages as a comparison object: It is a confirmed Ic-BL SN, was detected with NIRSpec (DDT programme 4554; PI: Martin-Carrillo) at an almost identical rest-frame phase \citep[27 days;][]{Martin-Carrillo_23lcr}, and at an almost identical redshift \citep[$z = 1.0272$;][]{Perley_23lcr}. Its spectrum\footnote{This spectrum was extracted from the single row of pixels that contains the SN for a more direct comparison with the 250702B spectra. A full analysis of SN\,2023lcr will be published in Martin-Carrillo et al. (2025, in prep).} is shown alongside the line-by-line pixel extractions (Section~\ref{sec:2d_spec}) in Figure~\ref{fig:pixelspec}. GRB\,250702B was covered by pixel row 32, though the PSF of $0.1$" closely matches the pixel size, so transient light may also be found in adjacent rows. It is immediately apparent that no bright SN resembling SN\,2023lcr is present in any of the spectra, though it is unclear whether one is absent or just extincted by dust along the line of sight.

To place limits on the presence of any supernova we must first attempt to remove the host galaxy light. Although the host spectral shape is similar across all wavelengths there are subtle differences. We therefore chose to use pixel 34 as a subtraction template, since it is the closest match to pixel 32 in the red (where SN contribution should be smallest). It is sufficiently far from the transient to have minimal contamination and is also the brightest pixel in the galaxy, hence providing the least additional noise in the subtraction. We scale the spectra to match in the 4-4.5 micron region and then subtract, with the resulting subtraction shown in Figure~\ref{fig:pixelspec}.

For comparison, we also extract the SN\,2023lcr spectrum row by row to obtain an appropriate comparison in flux space. In the case of SN\,2023lcr there is minimal host galaxy contribution and the peak pixel contains $\sim 80$\% of the total SN flux. We redden the SN\,2023lcr spectra by $A_V=3.1$ as indicated by the galaxy spectrum (Table~\ref{tab:per-pix-linefluxes}), and also by $A_V=5.8$ as the implied line-of-sight extinction to the burst (Section~\ref{sec:afterglow} and Figure~\ref{fig:afterglow-corner}), using a \citet{Cardelli89} extinction law with $R_V = 3.1$. Given that the extinction determination from the galaxy is set by the ensemble properties of the stellar population, we believe the direct line-of-sight extinction determined from the afterglow is more likely to be relevant. 

We note that the details of the subtraction limit are sensitive to the scaling of the host galaxy for subtraction. In particular, because of the heavy extinction, the spectral shape expected for the SN and for the stellar population is similar, especially at low signal-to-noise. Hence, some transient contribution cannot be excluded, although broad features comparable to SN\,2023lcr are not visible in the subtraction (see Figure~\ref{fig:pixelspec}). For the low extinction scenario the absence of these features is constraining, and we estimate that we would detect these features in subtractions to luminosities of 2-3 times fainter than AT2023lcr. However, since these are rest-frame optical/nIR features, they are substantially impacted by the additional dust in the higher extinction scenario, and we could not identify supernovae substantially fainter than AT2023lcr in this case. 

As an orthogonal test of our sensitivity to an SN\,2023lcr-like supernova, we subtract decreasing fractions of the SN\,2023lcr spectrum from the pixel 32 spectrum and measure whether the resulting flux deviations are greater or smaller than the measurement errors. This method is complementary to the host subtraction test because it essentially measures how faint a supernova could be recovered with a `perfect' host subtraction. We subtract the reddened SN\,2023lcr spectrum from pixel 32 (again using both $A_V = 3.1$ and $A_V = 5.8$), and divide the difference by the measurement uncertainties of the two original spectra added in quadrature.
For $A_V = 3.1$, we find that for an SN with flux half that of SN\,2023lcr, the subtraction residuals are consistently between 2 and 5 times the measurement errors in the region between 1\,$\mu$m and 3\,$\mu$m (observed), while an SN 5x fainter than SN\,2023lcr causes deviations between $1\sigma$ and $2\sigma$ in this interval. However, for the expected line-of-sight $A_V = 5.8$ from afterglow modelling, the subtraction residuals are at the $1\sigma$ level for a SN half the flux of SN\,2023lcr, and only reach $3\sigma$ for an equally bright SN. We therefore conclude that while we can exclude an SN of the same luminosity as SN\,2023lcr, we cannot exclude SNe much fainter than it.

Ultimately, a second epoch of spectroscopy may reveal any transient light. Imaging observations also have the potential to identify such signatures, although we caution that with the heavy extinction it may be extremely difficult to use the properties of any transient observed only photometrically to place meaningful progenitor constraints.

\subsection{Tidal disruption event limits}\label{sec:TDE}

A TDE from an SMBH was strongly disfavoured as the origin for GRB\,250702B based on its off-nuclear position. We also note that the relatively flat optical -- nIR SED of the jetted TDE AT2022cmc at an observed epoch of 41 - 46 days \citep{Pasham23} should produce a transient feature with extinction-corrected flux densities on the order of $\gtrsim 0.1$\,$\mu$Jy (for $A_V = 5.8$). While this SED was taken at $\sim 21$ rest frame days (compared to $\sim 25.5$ for GRB\,250702B), the light curve was relatively flat at this time (t$^{-0.3}$) and, at $z=1.193$, AT2022cmc was also marginally more distant. The absence of an additional emission component in pixel 32 and adjacent rows therefore provides further evidence against a jetted TDE similar to AT2022cmc, although the diversity of optical/nIR emission in this population is not well known.

The X-ray (0.5-4 keV) peak flux of GRB 250720B was measured by the Einstein Probe WXT instrument to be $(5.5\pm0.7)\times10$~erg~cm$^{-2}$~s$^{-1}$ (\citealt{2025GCN.40906....1C}). With our redshift measurement, this converts to a peak (0.5-4 keV) X-ray luminosity of $\mathrm{L_{X,peak}}=3\times10^{48}$~erg~s$^{-1}$. This means that any non-relativistic TDE scenario where the $\mathrm{L_{X,peak}}$ is not too far above the Eddington limit is ruled out (given that $\mathrm{L_{X,peak}}$ for even a black hole as massive as $10^8$~M$_\odot$ SMBH would be $\approx 2\times10^{46}$~erg~s$^{-1}$). Therefore, any TDE interpretation implies a strongly jetted (i.e., relativistic and Doppler-boosted) emission. This makes determining the mass from the BH responsible for the disruption uncertain, unlike for non-relativistic TDEs where the peak luminosity roughly scales with BH mass \citep{Hammerstein25}. The time scale of the repeated gamma-ray detection is consistent with the orbital time scale of a white dwarf around an IMBH. Even for rapidly spinning BHs, the maximum mass of a BH able to tidally disrupt outside the event horizon will be $\lessapprox 10^6$~M$_\odot$, thus, in the not well-probed IMBH regime \citep[see the review in][]{Maguire20}.

Interestingly, another high-energy event (albeit in X-ray and not gamma-ray) was reported in \citet{Jonker13}. In that case, two precursor events with time scales of $\approx$4000~s were reported, and a similar scenario of a white dwarf IMBH TDE was envisaged. However, for that source, no redshift determination was available, making the associated energy much more uncertain than for GRB\,250702B. Deep host galaxy searches \citep{Eappachen22} did provide evidence for redshift scenarios that could imply a peak X-ray luminosity consistent with what we find for GRB\,250702B.

\section{Host galaxy properties}\label{sec:galaxy}

\begin{table}
  \centering
  \caption{The properties of GRB\,250702B's host galaxy inferred from spectral fitting, as outlined in Section \ref{sec:galaxy}. Median a posteriori parameter values are given, with uncertainties bounding 68\% confidence intervals. Since the spectrum was scaled to match the host-integrated F160W {\em HST} measurement, the values below assume that the portion of the galaxy in the slit is representative of the stellar population throughout the host.}
  \renewcommand{\arraystretch}{1.4}
    \begin{tabular}{lc}
      \hline
      Parameter 
      & {\sc cigale} \\
      \hline
      $\log_{10}(M_*/M_\odot)$ 
      & $11.60^{+0.14}_{-0.14}$ \\
      $t_{\rm age}$ [Gyr] 
      & 4.9$\pm$0.6 \\ 
      $\tau$ [Gyr] 
      & $0.550^{+0.180}_{-0.180}$ \\ 
      $A_V$ [mag] 
      & 6.2$\pm$0.1 \\
      SFR [M$_{\odot}$/yr] 
      & 93$\pm$11 \\
      \hline
    \end{tabular}
  \label{tab:host_props}
\end{table}

\subsection{CIGALE results}

\begin{figure*}
\centering
    \includegraphics[width=16cm]{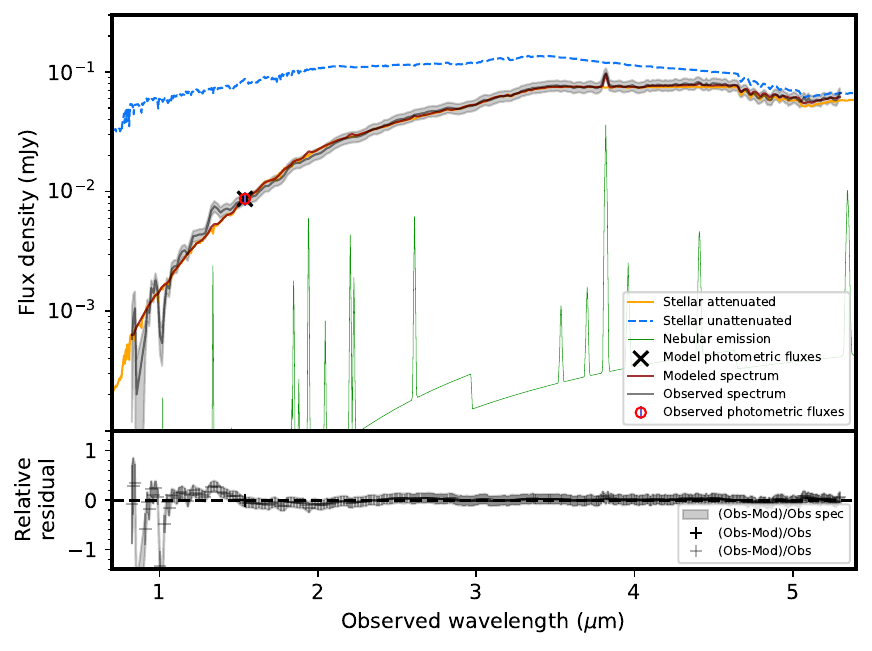}
    \caption{
    Best-fit model for the GRB 250702B host galaxy obtained with \textsc{Cigale}. The observed spectrum is shown as a gray curve, and the HST/WFC3 $F160W$  photometric point as a red circle. The resulting best-fit model is plotted in red, with the corresponding model photometric flux indicated by a black cross. The contributions from the attenuated stellar continuum (orange), intrinsic unattenuated stellar emission (blue), and nebular emission (green) are shown separately. The lower panel displays the relative residuals for both the spectrum (gray) and the photometric point (black). The inferred parameters are given in Table \ref{tab:host_props}.} 
    \label{fig:galfit}
\end{figure*}

We modeled the GRB\,250702B host galaxy using \textsc{Cigale} \citep{2019A&A...622A.103B,2025A&A...699A.336B}, in particular, its spectroscopic extension \textsc{Cigale-spec}, which can simultaneously fit 1D spectrum and broadband photometry. We adopt a delayed star formation history with a recent burst component, motivated by the recent star-formation activity typically observed in GRB host galaxies \citep{2009ApJ...691..182S,Kruhler15}. The age of the main stellar population is allowed to vary up to 5800 Myr, close to the maximum physically allowed at $z=1.036$. For the burst component, we explore ages between 50 and 200 Myr with mass fractions produced by the burst varying from 0 to 0.5. Stellar emission is modeled using the population synthesis models of \citet{2003MNRAS.344.1000B} with a Salpeter initial mass function \citep{Salpeter1955} and a metallicity set to $Z = 0.02$ (Solar metallicity). We included nebular emission with standard parameters and modeled dust attenuation using a modified version of the \citet{Calzetti2000} starburst attenuation law, with $A_V$ allowed to vary between 2.5 and 8.5 to enclose the value derived from emission lines and afterglow modeling. The resulting best-fit model is shown in Figure~\ref{fig:galfit}. The model reproduces both the continuum shape and the Pa$\alpha$ line flux well (model: $10.3\pm0.1\times10^{-18} \, \mathrm{erg}\,\mathrm{s}^{-1}\,\mathrm{cm}^{-2}$), but underestimates the H$\alpha$ flux (model: $1.96\pm0.20\times10^{-18} \, \mathrm{erg}\,\mathrm{s}^{-1}\,\mathrm{cm}^{-2}$) by a factor of $\sim5$. This discrepancy may reflect local variations in dust attenuation that suppress optical nebular emission more strongly than in the NIR and are not fully captured by the simple attenuation prescription adopted in our modeling. The Bayesian analysis indicates a main stellar population age of $(4.9\pm0.6)$ Gyr with an e-folding timescale of $\tau_\mathrm{main}=(550\pm180)$ Myr. A recent burst component is favored, with a characteristic age of $(129\pm46)$ Myr and a mass fraction of $f_\mathrm{burst}=0.050\pm0.004$, consistent with recent star formation activity. The model implies a dust attenuation corresponding to $E(B-V)_\mathrm{lines}=2.0\pm0.1$, i.e. a nebular $A_V = (6.2\pm0.1)$ mag, consistent with the afterglow-derived value. The host has a stellar mass of $\log_{10}(M_\star/M_\odot)=11.60\pm0.14$ and a star formation rate of $\log_{10}(\mathrm{SFR}/M_\odot\,\mathrm{yr}^{-1})=1.97\pm0.05$ ($\mathrm{SFR}=93\pm11~M_\odot\,\mathrm{yr}^{-1}$), placing it on or above the star forming main sequence at $z\approx1$ \citep{Schreiber15}.

\subsection{Star-formation rate from Balmer and Paschen lines}
We independently estimated the SFR from the observed Balmer and Paschen line fluxes using the equation scaled for solar metallicity in \cite{Reddy23}. After correcting the fluxes reported in Table~\ref{tab:lines} for host extinction with $A_V=3.1$~mag, we derive a $\mathrm{SFR}_{\mathrm{Pa}\alpha}=14.7\pm1.1~M_\odot\,\mathrm{yr}^{-1}$ and $\mathrm{SFR}_{\mathrm{H}\alpha}=7.0\pm1.1~M_\odot\,\mathrm{yr}^{-1}$. As Pa$\alpha$ is less affected by dust attenuation than H$\alpha$, its derived SFR is expected to exceed that from H$\alpha$. A factor up to $\sim2$ difference was observed in \cite{Reddy23}. The much larger SFR inferred from the SED fitting is explained by the normalization factor applied (approximately $\times10$), which is not used in the line-based estimates. Given that only a small fraction of the galaxy was covered by the NIRSpec slit and the high $A_V$ exhibited by the host, which may still affect the Pa$\alpha$ line, these SFRs should be considered as lower limits on the total SFR of the galaxy.

\subsection{Identified absorption features}

The CO first-overtone bandhead is reproduced within the 
\textsc{Cigale} code and this provides a direct perspective on the stellar population \citep[e.g.][]{Doyon1994,Origlia2000,Marmol-Queralto08}. In particular, this feature originates in cool stellar atmospheres, and can therefore arise both from red supergiants in starbursts galaxies, AGB stars at intermediate ages, and from the K/M-type giants that dominate older populations. The modest strength of the CO molecular band with $D_{CO}\sim1.15$ \citep[e.g. the fractional decrease of the continuum flux measured at the CO 2-0 absorption following the definition in][]{Marmol-Queralto08} suggests the integrated light is dominated by stars with effective temperatures of $\sim$4000\,K.
This interpretation is further strengthened by the likely presence of H$_2$O molecular absorption, which is also expected in the cool atmospheres of late-type stars \citep[e.g.][]{Lancon00}.

\section{Discussion}\label{sec:discussion}

\subsection{Rates and $E_{\gamma, \rm iso}$ in context}\label{sec:rates}
Combining our measured redshift of $z=1.036$ with the fluence measured by Konus-Wind \citep{Frederiks25}, we infer an isotropic-equivalent $\gamma$-ray energy release of $E_{\gamma, \rm iso} = (1.8 \pm 0.1) \times 10^{54}$\,erg (corresponding to a prompt efficiency of $\eta_\gamma\approx20\%$; Section~\ref{sec:afterglow}).
To properly compare this to GRBs at a range of redshifts, a cosmological k-correction must be applied to account for the shifting rest frame bandpass \citep[c.f.][]{Bloom01}. The best fit spectrum for the time-integrated Konus-Wind \citep{Aptekar95} observations was a power-law with a photon index of $\Gamma = 1.30 \pm 0.06$ in the energy range 20\,keV -- 1250\,keV. GRB $E_{\gamma, \rm iso}$ values are typically k-corrected to a bolometric rest-frame bandpass of 1\,keV -- 10\,MeV \citep[e.g.][]{Tsvetkova17,Tsvetkova21,Burns23}. This results in a k-corrected $E_{\gamma, \rm iso} = 8.2 \times 10^{54}$\,erg. This would make GRB\,250702B the second most energetic GRB of all time, behind only GRB\,221009A \citep[the `BOAT';][]{Burns23}. However, this is in part because the $\nu F_{\nu}$ peak of the spectrum is above the observed bandpass, making the energy solution unbound (i.e. tending to infinity as the bandpass increases). Assuming a peak energy of 1500\,keV (just above the measured energy range) gives $E_{\gamma, \rm iso} = 2.2 \times 10^{54}$\,erg, placing GRB\,250702B just below the 15 GRBs with $E_{\gamma, \rm iso} > 2.5 \times 10^{54}$\,erg presented in \citet{Burns23} in energy output. We note that the $E_{\gamma, \rm iso}$ calculated here is a lower limit, since related X-ray emission of an unreported duration was observed to emerge the day before the first \emph{Fermi}/GBM trigger by EP \citep{gcn40906}.

We also estimate the rarity of events like GRB\,250702B. Taking it as a singular event in the \emph{Fermi}/GBM catalogue, which began in 2009 \citep{Gruber14,vonKienlin14,Bhat16,vonKienlin20}, gives a time baseline of 16 years. To estimate the search volume, we take the least significant of the three sub-events, originally designated as GRB 250702D \citep{GRB250702D}, and calculate the maximum distance from which the burst could have been detected. The General Coordinates Network (GCN) notice associated with this event indicates that it was detected in flight with a significance of $5.1\sigma$ in the $4.096$\,s time bin. The \emph{Fermi}/GBM triggering algorithm requires a significance\footnote{\url{https://fermi.gsfc.nasa.gov/science/resources/swg/bwgreview/Trigger\_Algorithms.pdf}} of $4.5\sigma$ in two detectors, and hence with our measured distance of $7.1$\,Gpc, GRB 250702D could have been detected from a maximum distance of $D_{\rm max} = 7.1 / \sqrt{4.5/5.1} = 7.6$\,Gpc. The detection rate can therefore be approximated as $R_{\rm det} = 5 \times 10^{-4}$\,yr$^{-1}$\,Gpc$^{-3}$. Corrected for the very narrow jet found in afterglow modelling (Section~\ref{sec:afterglow}) and assuming a `top hat' jet, the true volumetric event rate is $R_{\rm evt} = R_{\rm det} / (1 - \cos(\theta_{\rm jet})) = 11$\,yr$^{-1}$\,Gpc$^{-3}$. This is approximately $10^{-5}$ times the estimated rate of core collapse SNe out to a similar redshift \citep{Dahlen12}, and before correction for beaming is a factor of $>10^3$ lower than the implied rate for high luminosity long GRBs \citep{sun15}.


 The above calculations are clearly simplified, neglecting occulted regions of the sky, variable sensitivity and background rates, and the impact redshifting has on received counts when observing with a fixed bandpass \citep[e.g.][]{Littlejohns13} among other factors. Nonetheless, with no comparable events in the archive, the true event rate is clearly low. We leave a full treatment of these factors to future work.

\subsection{Limits on the exclusion of an associated supernova}
To date we have observed more than 60 GRB-SN associations with almost half of them spectroscopically confirmed \citep[e.g.][]{Cano2017, Finneran25a}. The properties of these Ic-BL GRB-SNe are remarkably similar in all cases with similar spectroscopic features and evolutions \citep{Finneran25b}. Their light curves show narrow distributions in both the quasi-bolometric light-curve parameter $k_{\rm bol}$ and the quasi-bolometric stretch factor $s_{\rm bol}$ \citep[e.g.][]{Klose19}, signaling similar peak luminosities (within a factor of 3\,--\,5$\times$) with respect to the archetypal GRB-SN, SN\,1998bw, and similar progenitor stars. The derived limit suggests the possibility of an undetected SN component associated with GRB\,250702B with similar luminosity as SN\,2023lcr, which is, in terms of its peak magnitude, released mass of radioactive Nickel, kinetic energy and ejected mass, an  average case within the GRB-SN population \citep[e.g.][to see the range of values typically seen in GRB-SN associations for these parameters]{Cano2017, Finneran25a}. Thus, this limit only allows us to rule out the top $\sim$\,40\% most powerful GRB-SN cases, such as SN\,1998bw \citep[e.g.][]{Galama98}, SN\,2010ma \citep[e.g.][]{Sparre11}, SN\,2012bz \citep[e.g.][]{Schulze14} or SN\,2011kl \citep[e.g.][]{Gendre13} among others. Furthermore, if GRB-less Ic-BL SNe and other type of stripped enveloped SNe (such as Ic, Ib, IIb) are considered as potential associations with GRB\,250702B (pointing towards a different progenitor), the luminosity limit would be even less restricting than in the GRB-SN scenario \citep[e.g.][]{Lyman16, Afsariardchi21}, leaving these alternative progenitor scenarios still open.

\subsection{Galaxy context}

\begin{figure}
\centering
    \includegraphics[width=\columnwidth]{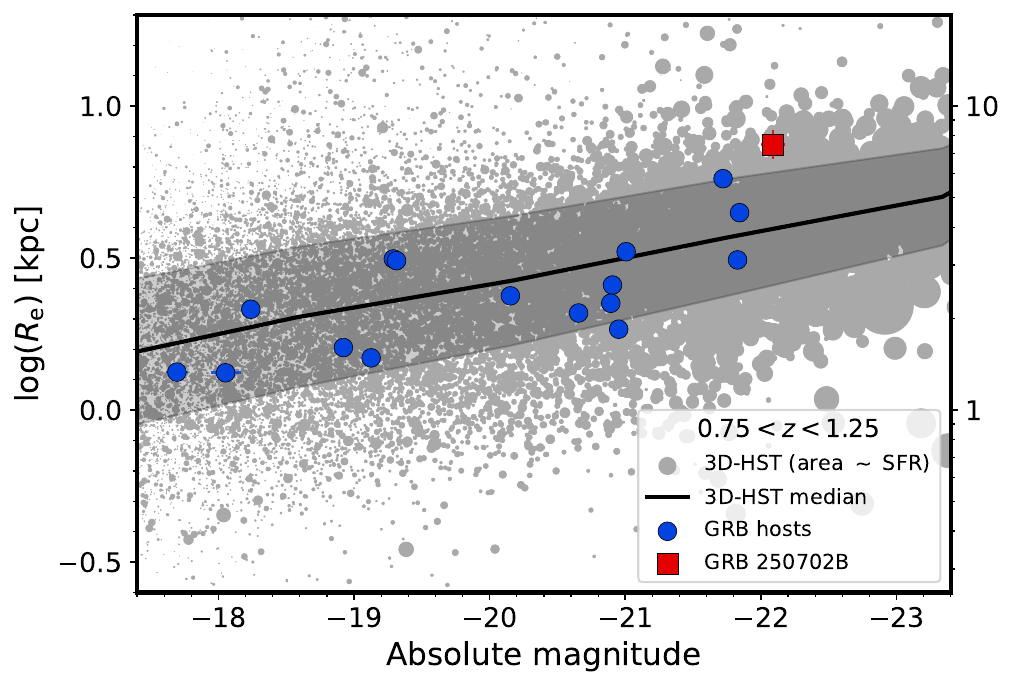}
    \includegraphics[width=\columnwidth]{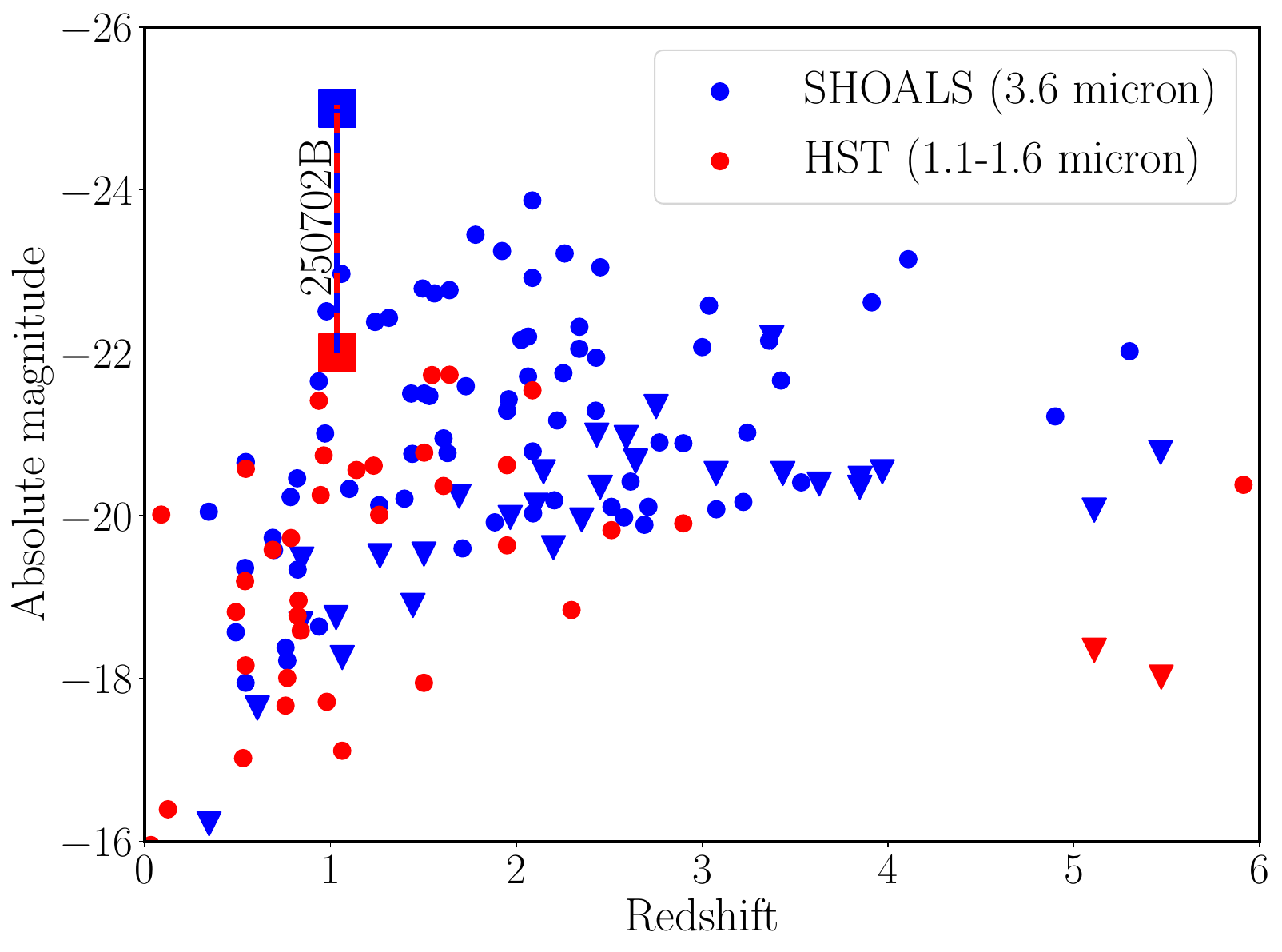}
    \caption{{\em Top:} Half-light radius ($R_\mathrm{e}$) as a function of $F160W$ ($\sim$H band) observed absolute magnitude for GRB host galaxies and 3D-HST star-forming galaxies at $0.75 < z < 1.25$. The red square marks the host galaxy of GRB\,250702B. Blue  circles show other GRB host galaxies, while gray circles represent star-forming galaxies, with their area scaled to the SFR. The black line indicates the median relation for the star-forming population, and the light gray shaded region shows its $1\sigma$ scatter. {\em Bottom:} Redshift vs absolute magnitude for GRB hosts across a wider redshift range. The host of GRB 250702B is the most luminous GRB host observed to date in both the near-IR and (likely) in the mid-IR.}
    \label{fig:mass_re}
\end{figure}

We compare the galaxy size (half-light radius, $R_\mathrm{e}$) and absolute magnitude of the GRB\,250702B host galaxy, as measured by \citet{Levan25}, with those of field star-forming galaxies at similar redshifts. 
Figure~\ref{fig:mass_re} shows the half-light radius as a function of observed absolute magnitude in the HST/WFC3 $F160W$ (approximately $H$-band) filter for GRB host galaxies from \cite{Blanchard16} and for star-forming galaxies drawn from the 3D-HST survey \citep{2012ApJS..200...13B,2014ApJS..214...24S,2014ApJ...788...28V,2016ApJS..225...27M} within $z \pm 0.25$. 
The host galaxy of GRB\,250702B (red square) lies significantly above the median size–luminosity relation of the field population and is located at the bright, large end of the star-forming galaxy distribution.
GRB host galaxies at these redshifts are typically star-forming, metal-poor, compact, and dense systems \citep{Fruchter06,Kelly14,Japelj16,Palmerio19,Schneider22}. More massive, dustier, and near- to super-solar metallicity hosts are rarer but have been reported \citep[e.g.,][]{2010ApJ...712L..26L,Schady15,perley16b}. The GRB\,250702B host stands out as one of the brightest and most extended known at this redshift range, indicating a relatively unusually massive and evolved system within the GRB host population.

In terms of morphology, the host has concentration and asymmetry \citep{conselice2005} values of $C=1.88$ and $A=0.059$ \citep[measured on the {\em HST} F160W image with {\sc statmorph},][]{statmoprh}. This is a typical asymmetry for GRB hosts. The concentration parameter, on the other hand, is among the lowest ever measured in the GRB host population \citep{lyman17,chrimes19}, and reflects the galaxy's highly extended nature outside the two prominent bright regions.

We further compare the properties of the host of GRB 250702B with the GRB host population across a wider range of redshift in Figure~\ref{fig:mass_re}. Comparison to the near-IR observations is undertaken directly with the {\em HST} integrated host measurement. To estimate the comparison to the much larger sample taken as part of the SHOALS survey \citep{perley16b}, we first estimate the colour based on that observed with NIRSPEC, and then apply the same normalisation from the NIRSPEC slit to the integrated light as used in the H-band. This is reasonable since there are not large colour gradients within our data, although there is a slight trend to redder colours nearer the galaxy nucleus. From this it is apparent that in both the near-IR (rest-frame) optical and mid-IR the host of GRB 250702B is the most luminous GRB host to date. Given the combination of exceptional burst and exceptional host it is tempting to speculate if the two are related.

ERIS imaging shows GRB\,250702B to lie in close proximity to structure in the host galaxy, which may well be related to the emission line structures seen in the JWST spectroscopy. This indicates the GRB did not come from the centre of mass of any large structure, although the emission lines are at their brightest at the burst location. The extended nature of these lines, and the fact that the burst does not lie at an apparent peak of the light within the host, may disfavour an IMBH TDE from the higher end of the IMBH mass function, which may be expected to dominate its local environment. The galaxy is apparently dominated by an old stellar population, which could imply an unusual channel. However, there is also evidence for significant star formation, in particular at the burst position, indicating that a supernova origin remains plausible. The very high mass of the galaxy in principle should yield a high metallicity, although robust metallicity estimates cannot be obtained from our spectroscopy. If this is the case, standard collapsar models may struggle to explain the burst, although more exotic scenarios may be possible, especially given the rarity of events like GRB\,250702B.

\section{Conclusions}\label{sec:conclusions}

We performed follow-up observations of the exceptional repeating GRB\,250702B with JWST/NIRSpec and VLT/ERIS. Our spectroscopic observations reveal the redshift of the GRB to be $z = 1.036 \pm 0.004$, which places a lower limit of $E_{\gamma \rm iso} > 2.2 \times 10^{54}$\,erg, assuming a conservative $\nu F_{\nu}$ peak energy of 1500\,keV and neglecting additional flux detected in X-rays by EP starting the day prior to the \emph{Fermi}/GBM triggers. This places GRB\,250702B among the very brightest GRBs in the observed catalog, though its extremely narrow jet opening angle inferred from afterglow modelling would result in more typical beaming-corrected GRB energies \citep[e.g.][]{Frail01}.

The host galaxy of GRB\,250702B is one of the intrinsically brightest and most spatially extended in the HST and SHOALS samples, particularly at 3.6\,$\mu$m, where it is almost a magnitude brighter than any other GRB host in absolute terms. The dominant stellar population in the region close to the GRB is found to be $4.9\pm0.6$\,Gyr old, although modelling with \textsc{Cigale} favors the inclusion of a burst component with a characteristic age of $129 \pm 46$\,Myr, indicating $\approx 5$\% of the stellar mass comes from recent star formation. The star formation rate itself is typical for a galaxy of this size and at this redshift. The galaxy is extremely dusty, with $A_V$ measurements ranging from $A_V = 3.1 \pm 0.2$ from Pa$\alpha$ to H$\alpha$ line ratios from the spectrum covering the GRB position, $A_V = 3.4 \pm 0.4$ from the same line ratios in the spatially integrated 1D NIRSpec spectrum, $A_V = 5.8 \pm 0.1$ along the GRB line-of-sight from afterglow modelling, to $A_V = 6.2 \pm 0.1$ from galaxy modelling.

Despite the extremely high visual extinction of the host, we are able to exclude the presence of a typical GRB-SN with a luminosity comparable to SN\,2023lcr. However, our observations are only constraining to the brightest 40\% of known GRB-SNe and are not sensitive to other SN types. We are therefore unable to exclude a stellar collapse origin for GRB\,250702B. We are also not able to constrain the presence of an IMBH TDE, although the observed position of the GRB is offset from the brightest regions of nearby star formation, indicating it did not come from the centre of a local mass structure as might be expected for BHs on the higher end of the IMBH mass spectrum.

Our observations have confirmed GRB\,250702B to be a surprisingly distant event given the observed brightness of its host galaxy. The associated energy release is enough to strain, but not definitely break, canonical GRB collapsar models. If it is a collapsar, we estimate it to be a 1 in 1000 event in the high-luminosity GRB population. The beaming-corrected event rate suggests it must come from a progenitor more than $10^5$ times rarer than core collapse supernovae. Ultimately, follow-up spectroscopic observations may help to reveal transient light in the spectra presented here that is obfuscated by the extremely dusty sight line to the burst, and elucidate the progenitor of this rare, powerful, and surprisingly distant relativistic transient.

\section{Acknowledgements}

This work is based in part on observations made with the NASA/ESA/CSA James Webb Space Telescope. The data were obtained from the Mikulski Archive for Space Telescopes at the Space Telescope Science Institute, which is operated by the Association of Universities for Research in Astronomy, Inc., under NASA contract NAS 5-03127 for JWST. These observations are associated with programs \#4554 and \#9254.
This work is based on observations made with the NASA/ESA Hubble Space Telescope.  These observations are associated with program 17988.  Based on observations collected at the European Southern Observatory under ESO programme(s) 114.27PZ.

This work is based in part on observations taken by the 3D-HST Treasury Program (GO 12177 and 12328) with the NASA/ESA HST, which is operated by the Association of Universities for Research in Astronomy, Inc., under NASA contract NAS5-26555.

BPG and Dimple acknowledge support from STFC grant No. ST/Y002253/1. BPG and DO acknowledge support from The Leverhulme Trust grant No. RPG-2024-117. BS acknowledges the support of the French Agence Nationale de la Recherche (ANR), under grant ANR-23-CE31-0011 (project PEGaSUS). AAC acknowledges support through the European Space Agency (ESA) research fellowship programme. ASn and DBM are funded by the European Union (ERC, HEAVYMETAL, 101071865). PGJ and MER are funded by the European Union (ERC, Starstruck, 101095973). Views and opinions expressed are however those of the authors only and do not necessarily reflect those of the European Union or the European Research Council Executive Agency. Neither the European Union nor the granting authority can be held responsible for them. AMC and LC acknowledge support from the Irish Research Council Postgraduate Scholarship No. GOIPG/2022/1008. The Cosmic Dawn Center (DAWN) is funded by the Danish National Research Foundation under grant DNRF140. LI acknowledges financial support from the INAF Data Grant Program `YES' (PI: Izzo) {\it Multi-wavelength and multi messenger analysis of relativistic supernovae}. GPL is supported by a Royal Society Dorothy Hodgkin Fellowship (grant Nos. DHF-R1-221175 and DHF-ERE-221005). ASa acknowledges support by a postdoctoral fellowship from the CNES. MEW is supported by the Science and Technology Facilities Council (STFC). 
NRT acknowledges support from STFC grant ST/W000857/1.

\bibliography{main}{}
\bibliographystyle{aasjournalv7}



\end{document}